\documentclass[12pt]{article}

\newcommand{\vr}{\mathbf{r}}
\newcommand{\vj}{\mathbf{j}}

\newcommand{\intep}{\mathbf{Z_+}}
\newcommand{\real}{\mathbf{R}}

\newcommand{\EQ}{\begin{equation}}
\newcommand{\EN}{\end{equation}}
\newcommand{\EQN}{\begin{eqnarray}}
\newcommand{\ENN}{\end{eqnarray}}
\newcommand{\CR}{\nonumber \\}

\newcommand{\EXP}{\exp}
\newcommand{\EX}{{\rm e}}
\newcommand{\I}{i}

\newcommand{\B}{\beta}
\newcommand{\D}{\delta}          

\newcommand{\G}{\gamma}
\newcommand{\GA}{\Gamma}

\newcommand{\E}{\epsilon}
\newcommand{\VE}{\varepsilon}

\newcommand{\HA}{{1 \over 2}}

\begin{document}
\newfont{\elevenmib}{cmmib10 scaled\magstephalf}
\renewcommand{\thefootnote}{\fnsymbol{footnote}}

\renewcommand{\thefootnote}{\arabic{footnote}}
\setcounter{footnote}{0}
\baselineskip=0.8cm

\begin{center}

\large

{\bf Statistical Mechanics for States with Complex Eigenvalues 

and

Quasi-stable Semiclassical Systems 

}

\end{center}

\begin{center}
\vskip20pt

T. KOBAYASHI$^1$ and T. SHIMBORI$^2$
\vskip3pt  
{\it  $^1$Department of General Education, Tsukuba College of Technology, 

Tsukuba, Ibaraki 305-0005, Japan

$^2$Institute of Physics, University of Tsukuba, Tsukuba, 
Ibaraki 305-8571, Japan}

E-mail: $^1$kobayash@a.tsukuba-tech.ac.jp 
and $^2$shimbori@het.ph.tsukuba.ac.jp
\vskip10pt

\end{center}

\begin{abstract}
{\small 
Statistical mechanics for states with complex eigenvalues, 
which are described by Gel'fand triplet and represent unstable states like 
resonances, are discussed 
on the basis of principle of equal ${\it a\ priori}$ probability. 
A new entropy corresponding to the freedom for the imaginary eigenvalues 
appears in the theory. 
In equilibriums it induces a new physical observable which can be identified 
as a common time scale. 
It is remarkable that in spaces with more than 2 dimensions 
we find out existence of stable and quasi-stable systems, 
even though all constituents are unstable. 
In such systems all constituents are connected by 
stationary flows which are generally observable and then 
we can say that they are semiclassical systems. 
Examples for such semiclassical systems 
are constructed in parabolic potential barriers. 
The flexible structure of the systems is  also pointed out. 

}

\end{abstract}

\hfil\break
{\bf 1. Introduction}
\vskip5pt

It is well known that quantum mechanics including complex eigenvalues 
which represent unstable states like resonances 
are described by Gel'fand triplet [1]. 
Since systems including such unstable states are usually unstable, 
it seems to be not suitable for constructing a theory like 
statistical mechanics for thermal equilibriums. 
We can, however, consider interesting situations, that is, 
the cases where the imaginary parts of energy eigenvalues of the states 
are so small that there are time enough to realize 
thermal equilibriums before the decay. 
In such situations statistical mechanics based on principle of equal 
${\it a\ priori}$ probability can be meaningful and 
we can observe some decay properties of such systems. 
It will possibly be a representation for quasi-stable systems having long 
life-times. 
Furthermore there is another noticeable feature of 
the complex eigenvalues such that they are 
represented by pairs of complex conjugates like $a\pm\I b$ for 
$a,b\in \real
$,
that is, since Hamiltonians $\hat H$ are real, for any solutions $\psi$ 
satisfying 
the equation 
$$
\hat H\psi=(a-\I b)\psi \ \ \ \ \ \ {\rm for} \ a,b \in \real,
$$ 
we always find solutions having complex conjugate eigenvalues such that 
$$
\hat H\psi^*=(a+\I b)\psi^*.
$$ 
An explicit example was presented for parabolic potential barrier [2,3,4]. 
It is quite interesting that in spaces with more than 2 dimensions 
real eigenvalue solutions can be obtained by using those pair solutions 
represented only by Gel'fand triplet and they are in 
infinite degrees of degeneracy [4]. 
In general they describe stationary states but they simultaneously 
have incoming- and outgoing-quantum 
stationary flows [4,5]. 
Actually 
the real eigenvalue solutions are interpreted as the states which have equal
incoming- and outgoing-flows. 
It is quite natural to have questions; whether 
systems constructed with such real-eigenvalue states can really be stable 
and furthermore 
how they can physically be interpreted in realistic phenomena. 
In order to study these problems 
we shall make statistical mechanics for states including 
complex eigenvalues on the basis 
of the same principle for the usual statistical mechanics, 
that is, principle of equal ${\it a\ priori}$ probability. 

In $\S$2 statistical mechanics for complex eigenvalue states is performed. 
It is shown that the entropy of a system is 
represented by the sum of the usual entropy  and a 
new one induced from the freedom of imaginary part and a new observable 
quantity appears in equilibriums, which will be identified 
as time scale of decay of the system. 
Explicit examples for stable and quasi-stable systems are presented 
in the case of parabolic potential barriers in $\S$3, 
where those quasi-stable systems are connected by stationary flows [5]. 
Some realistic applications will be remarked in $\S$4. 
\vskip50pt

\hfil\break
{\bf 2. Statistical mechanics for complex-eigenvalue states}
\vskip5pt

Let us construct a statistical mechanics corresponding to microcanonical 
ensemble for states having complex energies 
which are generally represented by 
\EQ
\VE_{ij_i}=\E_i-\I\G_{j_i}\ \ \ \ \ \ {\rm for}\ \E_i,\G_{j_i}
\in \real
\EN 
where $i,\ j_i\in \intep\ (\intep=\{0,1,2,...\}) 
$ 
and the suffix $i$ of $j_{i}$ is needed when there is some relation between 
the real and imaginary energy eigenvalues. 
We consider a simple case described by a system composed of 
$N$ independent particles being in complex-energy states. 
In this case 
the total energy of the $N$-particle system is given 
by the sum of energy eigenvalues of each particles such that 
\EQ
{\cal E}=E-\I\GA,
\EN 
where 
\EQ
E=\sum_i\E_i\ \ {\rm and}\ \ \GA=\sum_{j_i}\G_{j_i}.
\EN 
Here we shall investigate simple cases where the real and imaginary energy 
eigenvalues are independently determined and then 
we can take off the suffix $i$ from $j_i$. 
Such models will explicitly be presented in the next section. 
The basic principle is taken as same as 
that for the usual statistical mechanics, 
that is, principle of equal ${\it a\ priori}$ probability. 
Then we start from  counting the number of independent combinations of states
for a fixed energy ${\cal E}$. 
Since two freedoms concerning to the real and imaginary parts of energies are 
independent of each other, 
the number of the combination $W({\cal E})$ is counted by the product of 
the number $W^\Re(E)$ for realizing the real part $E$ and that 
$W^\Im(\GA)$ for realizing the imaginary part $\GA$ such that
\EQ
W({\cal E})=W^\Re(E) W^\Im(\GA).
\EN 
Following the procedure of statistical mechanics, 
we now see that 
the entropy $S({\cal E})=k_\mathrm{B}\log W({\cal E})$ 
of the system is written 
in terms of the sum of two entropies 
such that 
\EQ
S({\cal E})=S^\Re(E)+S^\Im(\GA),
\EN 
where 
$S^\Re(E)=k_\mathrm{B}\log W^\Re(E)$ and 
$S^\Im(\GA)=k_\mathrm{B}\log W^\Im(\GA)$ 
are, respectively, 
the Boltzmann entropy  and the new entropy induced from the freedom of 
the imaginary part. 

Let us consider equilibrium between two systems which can each other 
transfer only energies. 
The total energy ${\cal E}=E-\I\GA$ given by the sum of those for 
two systems  ${\cal E}_{\rm I}=E_{\rm I}-\I \GA_{\rm I}$ and 
${\cal E}_{\rm II}=E_{\rm II}-\I \GA_{\rm II}$ is fixed. 
The number of the available combinations is written by the product of those 
for the two systems as
\EQ
W({\cal E})=W_{\rm I}({\cal E}_{\rm I}) W_{\rm II}({\cal E}_{\rm II}),
\EN 
where $W_{\rm I}({\cal E}_{\rm I})
=W_{\rm I}^\Re(E_{\rm I}) W_{\rm I}^\Im(\GA_{\rm I})$ and 
$W_{\rm II}({\cal E}_{\rm II})
=W_{\rm II}^\Re(E_{\rm II}) W_{\rm II}^\Im(\GA_{\rm II})$. 
Now we have the entropy expressed as the sum of four terms 
\EQ
S({\cal E})=S_{\rm I}^\Re(E_{\rm I})+S_{\rm I}^\Im(\GA_{\rm I})+
S_{\rm II}^\Re(E_{\rm II})+S_{\rm II}^\Im(\GA_{\rm II}),
\EN 
where $S_{\rm I}^\Re(E_{\rm I})
=k_\mathrm{B}\log W_{\rm I}^\Re(E_{\rm I})$ and so on. 
In the procedure maximizing the entropy $S({\cal E})$ 
under the constraints that $E=E_{\rm I}+E_{\rm II}$ and 
$\GA=\GA_{\rm I}+\GA_{\rm II}$ are fixed, 
we obtain two independent relations corresponding to 
the two constraints such that 
\EQ
{\partial S_{\rm I}^\Re(E_{\rm I}) \over \partial E_{\rm I}}=
{\partial S_{\rm II}^\Re(E_{\rm II}) \over \partial E_{\rm II}},
\EN 
\EQ
  {\partial S_{\rm I}^\Im(\GA_{\rm I}) \over \partial \GA_{\rm I}}=
{\partial S_{\rm II}^\Im(\GA_{\rm II}) \over \partial \GA_{\rm II}}.
\EN 
The first relation leads the usual temperature but the second one produce 
a new quantity which must be same for the two systems in the equilibriums. 
The canonical distribution for energy 
${\cal E}_{lm}=E_l-\I \GA_m$ is written by 
\EQ
P({\cal E}_{lm})=Z^{-1}\EXP(-\B^\Re E_l-\B^\Im\GA_m),
\EN 
where $\B^\Re$ should be chosen as the usual factor 
$\B=(k_\mathrm{B}T)^{-1}$ of 
canonical distribution, 
$\B^\Im$ denotes the new physical quantity in the equilibriums and 
the canonical partition function $Z$ is given by 
$$
Z=\sum_l \sum_m \EXP(-\B E_l-\B^\Im\GA_m).
$$ 
Note that we can not derive the canonical distribution (10) 
by replacing a real energy $E$ with a complex energy ${\cal E}$ 
in the usual canonical distribution 
$
P(E)=Z^{-1}\exp(-\B E)
$. 

What is the new quantity $\B^\Im$? 
In independent particle systems wave functions are written by the product 
of all constituents such that 
\EQ 
\Psi(t,\vr_1,...,\vr_N|{\cal E})=\prod_{n=1}^N 
\psi(t,\vr_n|\VE_n),
\EN 
where the wave function for one constituent with 
$\VE_n=\E_n-i\G_n$ is generally given by 
\EQ
\psi(t,\vr_n|\VE_n)=\EX^{-\I \VE_n t/\hbar}\phi(\vr_n).
\EN 
The probability density for $\Psi$ at $t$, which is normalized 
at $t=0$, is evaluated as 
\EQN
\rho(t,\vr_1,...,\vr_N|{\cal E})
&=&|\Psi(t,\vr_1,...,\vr_N|{\cal E})|^2 \CR
    &=&\prod_n \EX^{-2\G_n t/\hbar} |\phi(\vr_n)|^2 \CR
    &=&\EX^{-2\GA t/\hbar}\prod_n|\phi(\vr_n)|^2 .
\ENN 
We see 
that all the states with the same total imaginary energy $\GA$ have the same 
time-dependence $\EX^{-2\GA t/\hbar}$. 
Since the states with complex energy eigenvalues are unstable, 
the canonical distribution $P({\cal E})$ must depend on time. 
It is natural that the time-dependence of $P({\cal E})$ is same as that of 
the probability density, which is determined by the imaginary part $\GA$ 
of the total energy ${\cal E}$ of the system. 
We can specify 
\EQ
\B^\Im=2t/\hbar.
\EN 
Thus we can introduce 
a common time scale $t$ for two systems being in equilibriums. 

Remember 
that the imaginary parts $\G_j$ are expressed by pairs of conjugates, 
that is, $\pm |\G_j|\ \ (\forall j \in \intep)$ as noted in $\S$1. 
This fact means that the total imaginary part $\GA$ can possibly be in 
microscopic order 
(quantum size), even if the total real part $E$ is in macroscopic order. 
In fact such situations can be realized, 
when most of the constituents are 
in pairs of $\pm |\G_j|$ and number of non-pairing particles are 
very small in comparison with the macroscopic number $N$. 
In special cases $\GA=0$ can happen. 
These systems have real energies and then there is no reason for the systems 
to be unstable in time. 
Remembering that the states with positive imaginary parts 
and those with negative ones, respectively, 
represent growing- and decaying-resonance states [1,2], 
we see that in such systems two processes, that is, growing- and decaying-
resonance processes occur with the same probability. 
It is also noticeable that such systems with real energies are in 
infinite degrees of degeneracy, because the combinations of the pairing 
with zero imaginary energy are infinity. 
We will definitely see this situation 
in parabolic-potential-barrier models in the next section. 
\vskip10pt

\hfil\break
{\bf 3. Stable and quasi-stable systems connected by quantum stationary flows}

\vskip5pt
In the previous section the existence of stable many resonance systems have 
been pointed out. 
How can we understand such systems? 
Let us start from studying an explicit example 
in the 2-dimensinal parabolic potential case 
discussed in the paper [4]. 
Before discussing on the imaginary energies, we note that the real energy 
eigenvalue is always represented by a fixed constant in parabolic potentials 
[2,3] 
and then there is no freedom arising from the real energy eigenvalue.  
Imaginary eigenvalues $\G_j$ for the potential are given by the sum of two 
eigenvalues for two independent 1-dimensional parabolic potentials 
($V(x,y)=-{1 \over 2}m\G^2(x^2+y^2)$) such that 
\EQ 
\G_{j_1j_2}=j\hbar\G, \EN 
where $j=j_1+j_2$ with $j_1=\pm(n_1+ {1 \over 2})$ and $j_2=\pm(n_2+ 
{1 \over 2})$ for $n_{1},n_2\in \intep$. 
It is interesting that we have $\G_{j_1j_2}=0$ for the cases with $j_1=-j_2
=\pm(n+{1 \over 2})\ \ (\forall n\in \intep)$. 
Note that the zero imaginary eigenvalues are infinitely degenerate. 
As discussed in ref.~4, such zero imaginary states are well interpreted by 
the figure 
in fig.~1, where they are described by the wave function 
$(\psi_{n_1n_2} (t,\vr)) $ 
with equal incoming- and 
outgoing-stationary flows which are defined by the probability current 
$$
\vj(t,\vr)=\Re [\psi_{n_1n_2} (t,\vr)^*(-\I\hbar\nabla) 
\psi_{n_1n_2} (t,\vr)]/m
$$ 
with $m=$the mass of particle [4,5]. 

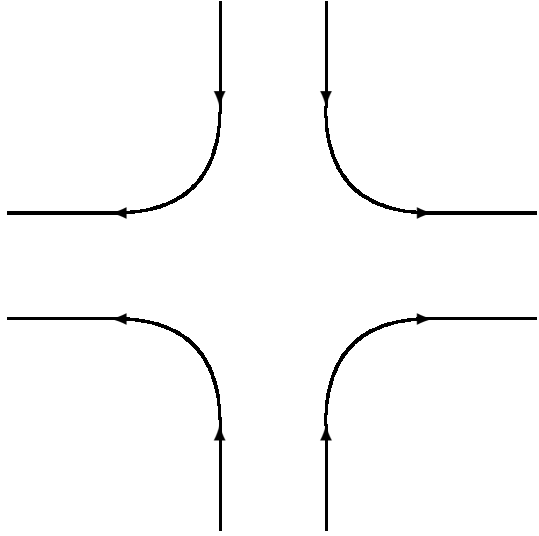
\begin{figure}[thbp]
\thicklines
\begin{center}
  \begin{picture}(200,200)
   
   \put(120.5,200){\vector(0,-1){40}}
   \qbezier(120,160)(120,120)(160,120)
   \put(160,120){\vector(1,0){1}}
   \put(160,120){\line(1,0){40}}
   
   \put(80.4,200){\vector(0,-1){40}}
   \qbezier(80,160)(80,120)(40,120)
   \put(40,120){\vector(-1,0){1}}
   \put(40,120){\line(-1,0){40}}
   
   \put(80.4,0){\vector(0,1){40}}
   \qbezier(80,40)(80,80)(40,80)
   \put(40,80){\vector(-1,0){1}}
   \put(40,80){\line(-1,0){40}}
   
   \put(120.5,0){\vector(0,1){40}}
   \qbezier(120,40)(120,80)(160,80)
   \put(160,80){\vector(1,0){1}}
   \put(160,80){\line(1,0){40}}
  \end{picture}
\end{center}
\caption[]{Stationary flow 
in a 2-dimensional parabolic potential, 
of which center is at the center of the figure. }
\end{figure}

We see that 
the stationary flow incoming toward the center of the potential 
is expressed by the positive eigenvalue and that outgoing from the center done 
by the negative one. (In details, see ref.~4.)  

Let us here consider $N$ pairs of a particle and a parabolic potential 
arranged on a line. 
Since the magnitudes of the two flows are equal at the same distance from the 
center, we can connect the flow outgoing from a potential to 
that incoming toward the next potential at the middle point of 
the two potentials.  
Such a situation is expressed in fig.~2. 

\begin{figure}[htbp]
  \thicklines
\begin{center}
\begin{picture}(420,160)
   
   \put(352.5,160){\vector(0,-1){32}}
   \qbezier(352,128)(352,96)(384,96)
   \put(384,96){\vector(1,0){1}}
   \put(384,96){\line(1,0){32}}
   
   \put(320.5,160){\vector(0,-1){32}}
   \qbezier(320,128)(320,96)(288,96)
   \put(288,96){\vector(-1,0){16}}
   \put(272,96){\line(-1,0){16}}
   \qbezier(256,96)(224,96)(224,128)
   \put(224.5,128){\vector(0,1){1}}
   \put(224.5,128){\line(0,1){32}}

   \put(96.5,160){\vector(0,-1){32}}
   \qbezier(96,128)(96,96)(128,96)
   \put(128,96){\vector(1,0){16}}
   \put(144,96){\line(1,0){16}}
   \qbezier(160,96)(192,96)(192,128)
   \put(192.5,128){\vector(0,1){1}}
   \put(192.5,128){\line(0,1){32}}

   \put(64.5,160){\vector(0,-1){32}}
   \qbezier(64,128)(64,96)(32,96)
   \put(32,96){\vector(-1,0){1}}
   \put(32,96){\line(-1,0){32}}
   
   \put(64.5,0){\vector(0,1){32}}
   \qbezier(64,32)(64,64)(32,64)
   \put(32,64){\vector(-1,0){1}}
   \put(32,64){\line(-1,0){32}}

   \put(96.5,0){\vector(0,1){32}}
   \qbezier(96,32)(96,64)(128,64)
   \put(128,64){\vector(1,0){16}}
   \put(144,64){\line(1,0){16}}
   \qbezier(160,64)(192,64)(192,32)
   \put(192.5,32){\vector(0,-1){1}}
   \put(192.5,32){\line(0,-1){32}}

   \put(320.5,0){\vector(0,1){32}}
   \qbezier(320,32)(320,64)(288,64)
   \put(288,64){\vector(-1,0){16}}
   \put(272,64){\line(-1,0){16}}
   \qbezier(256,64)(224,64)(224,32)
   \put(224.5,32){\vector(0,-1){1}}
   \put(224.5,32){\line(0,-1){32}}
   
   \put(352.5,0){\vector(0,1){32}}
   \qbezier(352,32)(352,64)(384,64)
   \put(384,64){\vector(1,0){1}}
   \put(384,64){\line(1,0){32}}
  \end{picture}
\end{center}
\caption[]{1-dimensional lattice connected by 
stationary flows. }
\end{figure}
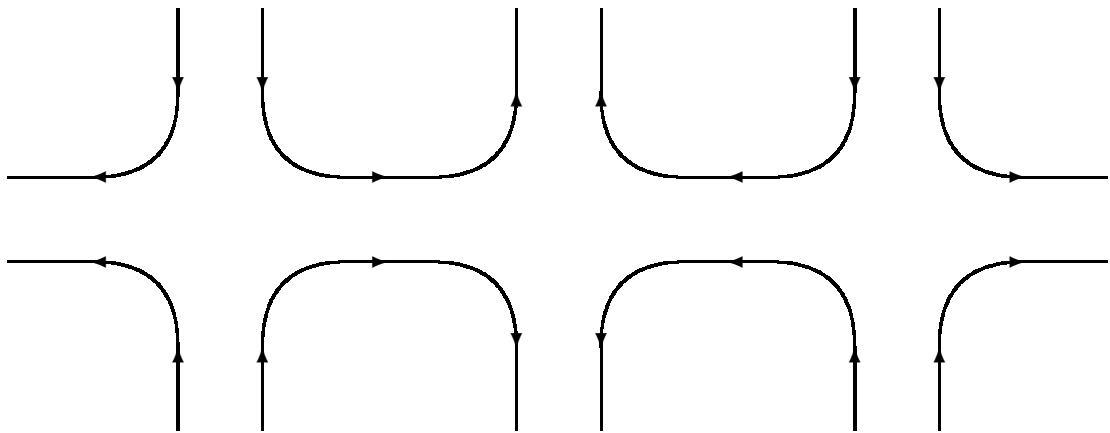

Now we can easily construct 2-dimensional lattices 
in which all lattice points are connected 
by stationary flows as figured in fig.~3.
It is remarkable that 
we can have stable systems composed of unstable constituents such as 
resonances. 
Note also that every 
\begin{figure}[tbph]
\thicklines
\begin{center}
  \begin{picture}(320,320)
    
   \put(264.5,312){\vector(0,-1){24}}
   \qbezier(264,288)(264,264)(288,264)
   \put(288,264){\vector(1,0){1}}
   \put(288,264){\line(1,0){24}}

   \put(240.5,312){\vector(0,-1){24}}
   \qbezier(240,288)(240,264)(216,264)
   \put(216,264){\vector(-1,0){12}}
   \put(204,264){\line(-1,0){12}}
   \qbezier(192,264)(168,264)(168,288)
   \put(168.5,288){\vector(0,1){1}}
   \put(168.5,288){\line(0,1){24}}

   \qbezier(168,192)(168,168)(192,168)
   \put(192,168){\vector(1,0){12}}
   \put(204,168){\line(1,0){12}}
   \qbezier(216,168)(240,168)(240,192)
   \put(240.5,192){\vector(0,1){12}}
   \put(240.5,204){\line(0,1){12}}
   \qbezier(240,216)(240,240)(216,240)
   \put(216,240){\vector(-1,0){12}}
   \put(204,240){\line(-1,0){12}}
   \qbezier(192,240)(168,240)(168,216)
   \put(168.5,216){\vector(0,-1){12}}
   \put(168.5,204){\line(0,-1){12}}

   \put(312,168){\vector(-1,0){24}}
   \qbezier(288,168)(264,168)(264,192)
   \put(264.5,192){\vector(0,1){12}}
   \put(264.5,204){\line(0,1){12}}
   \qbezier(264,216)(264,240)(288,240)
   \put(288,240){\vector(1,0){1}}
   \put(288,240){\line(1,0){24}}

   \put(72.5,312){\vector(0,-1){24}}
   \qbezier(72,288)(72,264)(96,264)
   \put(96,264){\vector(1,0){12}}
   \put(108,264){\line(1,0){12}}
   \qbezier(120,264)(144,264)(144,288)
   \put(144.5,288){\vector(0,1){1}}
   \put(144.5,288){\line(0,1){24}}
   
   \put(48.5,312){\vector(0,-1){24}}
   \qbezier(48,288)(48,264)(24,264)
   \put(24,264){\vector(-1,0){1}}
   \put(24,264){\line(-1,0){24}}
  
   \put(0,168){\vector(1,0){24}}
   \qbezier(24,168)(48,168)(48,192)
   \put(48.5,192){\vector(0,1){12}}
   \put(48.5,204){\line(0,1){12}}
   \qbezier(48,216)(48,240)(24,240)
   \put(24,240){\vector(-1,0){1}}
   \put(24,240){\line(-1,0){24}}

   \qbezier(144,192)(144,168)(120,168)
   \put(120,168){\vector(-1,0){12}}
   \put(108,168){\line(-1,0){12}}
   \qbezier(96,168)(72,168)(72,192)
   \put(72.5,192){\vector(0,1){12}}
   \put(72.5,204){\line(0,1){12}}
   \qbezier(72,216)(72,240)(96,240)
   \put(96,240){\vector(1,0){12}}
   \put(108,240){\line(1,0){12}}
   \qbezier(120,240)(144,240)(144,216)
   \put(144.5,216){\vector(0,-1){12}}
   \put(144.5,204){\line(0,-1){12}}

   \qbezier(144,120)(144,144)(120,144)
   \put(120,144){\vector(-1,0){12}}
   \put(108,144){\line(-1,0){12}}
   \qbezier(96,144)(72,144)(72,120)
   \put(72.5,120){\vector(0,-1){12}}
   \put(72.5,108){\line(0,-1){12}}
   \qbezier(72,96)(72,72)(96,72)
   \put(96,72){\vector(1,0){12}}
   \put(108,72){\line(1,0){12}}
   \qbezier(120,72)(144,72)(144,96)
   \put(144.5,96){\vector(0,1){12}}
   \put(144.5,108){\line(0,1){12}}

   \put(0,144){\vector(1,0){24}}
   \qbezier(24,144)(48,144)(48,120)
   \put(48.5,120){\vector(0,-1){12}}
   \put(48.5,108){\line(0,-1){12}}
   \qbezier(48,96)(48,72)(24,72)
   \put(24,72){\vector(-1,0){1}}
   \put(24,72){\line(-1,0){24}}

   \put(48.5,0){\vector(0,1){24}}
   \qbezier(48,24)(48,48)(24,48)
   \put(24,48){\vector(-1,0){1}}
   \put(24,48){\line(-1,0){24}}

   \put(72.5,0){\vector(0,1){24}}
   \qbezier(72,24)(72,48)(96,48)
   \put(96,48){\vector(1,0){12}}
   \put(108,48){\line(1,0){12}}
   \qbezier(120,48)(144,48)(144,24)
   \put(144.5,24){\vector(0,-1){1}}
   \put(144.5,24){\line(0,-1){24}}

   \put(312,144){\vector(-1,0){24}}
   \qbezier(288,144)(264,144)(264,120)
   \put(264.5,120){\vector(0,-1){12}}
   \put(264.5,108){\line(0,-1){12}}
   \qbezier(264,96)(264,72)(288,72)
   \put(288,72){\vector(1,0){1}}
   \put(288,72){\line(1,0){24}}
   
   \qbezier(168,120)(168,144)(192,144)
   \put(192,144){\vector(1,0){12}}
   \put(204,144){\line(1,0){12}}
   \qbezier(216,144)(240,144)(240,120)
   \put(240.5,120){\vector(0,-1){12}}
   \put(240.5,108){\line(0,-1){12}}
   \qbezier(240,96)(240,72)(216,72)
   \put(216,72){\vector(-1,0){12}}
   \put(204,72){\line(-1,0){12}}
   \qbezier(192,72)(168,72)(168,96)
   \put(168.5,96){\vector(0,1){12}}
   \put(168.5,108){\line(0,1){12}}

   \put(240.5,0){\vector(0,1){24}}
   \qbezier(240,24)(240,48)(216,48)
   \put(216,48){\vector(-1,0){12}}
   \put(204,48){\line(-1,0){12}}
   \qbezier(192,48)(168,48)(168,24)
   \put(168.5,24){\vector(0,-1){1}}
   \put(168.5,24){\line(0,-1){24}}
   
   \put(264.5,0){\vector(0,1){24}}
   \qbezier(264,24)(264,48)(288,48)
   \put(288,48){\vector(1,0){1}}
   \put(288,48){\line(1,0){24}}
  \end{picture}
\end{center}
\caption[]{2-dimensional lattice connected by 
stationary flows. }
\end{figure}
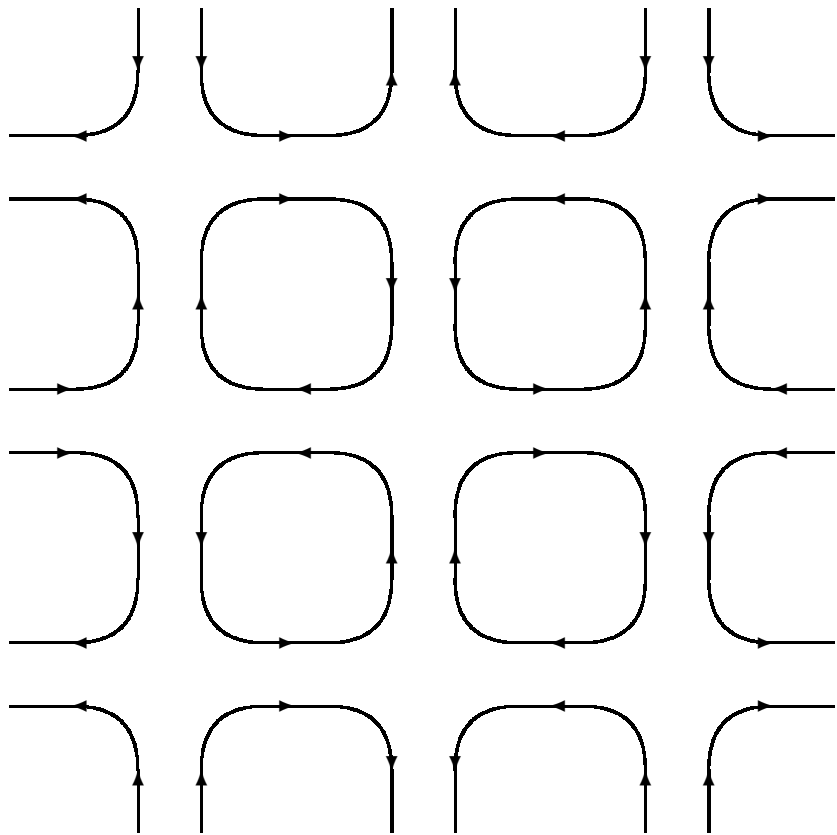
square made up from four lattice points is 
connected by a circular flow, that is, the total lattice system seems to be 
composed of small circular flows trapped in the squares. 
The fact that these lattices are not connected  
by wave functions but by stationary flows  means that 
they can have some classical properties, because the flows are basically 
observable in quantum mechanics. 
Therefore we can call the lattices semiclassical systems. 
Note here that length between two neighboring lattice points are 
arbitrary and then structures of lattices are not rigid but rather flexible. 

The following point should be noticed that at the edge of the system 
incoming- and outgoing-flows can generally not be connected, 
that is, the system have absorbing points and sending-out points 
of the stationary flows in  environments (heat baths) like breathing of 
living matters. 
This fact implies that those systems can generally be unstable 
and decay at the edge in the 2-dimensions 
when they are put in unsuitable environments, where 
the absorbing flows and the sending-out flows are not arranged with an equal
weight. 
Such quasi-stable systems, however, 
seems to be quite interesting to describe realistic 
systems slowly decaying from their edges, 
which can be seen quite often in our daily life. 
Of course, we can have stable systems on closed surfaces 
like soccer balls and torus. 
\vskip5pt

In 3-dimensional isotropic parabolic potentials we have 
no stable states with zero imaginary  energy eigenvalue, 
because energies induced from zero-point energy eigenvalues 
$\pm {\I \over 2}\hbar \G$ do not cancel each other in odd dimensions. 
Provided that the product of the curvature of potential $\G$ and 
the number of constituents $N$ is so small that the time scale given 
by $\tau=({1 \over 2}\G N)^{-1}$ is enough large 
in comparison with the time scale 
realizing thermal equilibriums, 
we can observe metastable systems which decay not only  on the edges 
but in the inside of the systems as well. 

Even in odd dimensions 
we find out two possibility that 
real eigenvalues appear from the compositions  
of complex eigenvalues in 3-dimensional spaces. 
One is the case where three coupling constants ($\G_1,\ \G_2,\ \G_3$) 
representing the curvatures in 3-dimensions 
are different from each other but satisfied by some relation, for instance, 
$
\G_1+\G_2=\G_3
$ 
and three eigenvalues ($\pm \I(n_1+\HA)\hbar\G_1,
\ \pm \I(n_2+\HA)\hbar\G_2,
\ \mp \I(n_3+\HA)\hbar\G_3$) fulfill the relations 
$
n_1\G_1+n_2\G_2=n_3\G_3.
$
States satisfying these two relations can have zero imaginary eigenvalues. 
Note that at least one state specified by $n_1=n_2=n_3=0$ satisfies 
the relations. 
These relations means that all flows incoming from two directions go out 
from one direction and {\it vice versa}. 
Another possibility for zero imaginary eigenvalues is seen in the case that 
potentials is described by 2-dimensional parabolic potential 
and 1-dimensional harmonic one. 
Imaginary energy-eigenvalues of total systems 
can be zero. 
In these cases we can find very rich and complicated structures in 
quasi-stable systems. 
In both cases the decays of those systems can occur only 
on their surfaces including edges. 
\vskip1pt

\hfil\break
{\bf 4. Remarks}
\vskip5pt

We see that statistical mechanics based on principle of equal {\it a priori} 
probability can be applicable to systems composed of many unstable states. 
To complete statistical mechanics, we still have a lot of problems to 
investigate, for instance, how to define and interpret 
thermal quantities corresponding to 
free energy, pressure, heat capacity and so forth. 
Ergodic theorem have also to be discussed in the prsent scheme. 
We have, however, found out two remarkable results. 
One is the introduction of a common time scale, which will possibly have 
some connection with the common time of our universe. 
The other is the existence of stable and 
quasi-stable systems which seem to be interesting candidates 
to describe realistic systems having long life-times. 
It should be stressed that the structure of those systems can be very 
much flexible because the distances between any neighboring constituents 
are arbitrary. 

A toy model can represented by systems for gases of which constituents have 
parabolic potential barriers in each other. 
Such systems can completely be separable into center of mass motions and 
relative motions written by the same parabolic potential. 
The procedure for the statistical mechanics performed in $\S$2 
can directly be applicable to this model and reproduce usual temperature 
from the center of mass motions and time scale from relative motions. 
In this model growing- and decaying- resonance states, respectively, describe 
two constituents approaching and leaving each other. 
Actually 2-dimensional models seems to be applicable to  the investigation 
of problems with respect to surfaces of materials. 

It is noticeable that many body systems can possibly be quasi-stable 
even under such repulsive potentials. 
Systems composed of many bodies having repulsive forces between any pairs of 
constituents 
will be realized electron gases, that is, electron plasma. 
We have to solve complex eigenvalue problem 
for the repulsive Coulomb potential. 
In order to say whether such quasi-stable systems exist or not, 
it is enough to show whether there exist complex eigenvalue solutions or not. 
We need not understand any details of the complex eigenvalues, 
because the existence of complex conjugate pairs is guaranteed in the theory. 
We shall study this problem in more detail. 

Up to now we have discussed on repulsive potentials.  
As studied in textbooks [6], it is known that 
attractive potentials such as shown in fig.~4 have complex eigenvalue 
solutions. 
\begin{figure}[htbp]
\thicklines
\vskip60pt
\begin{center}
  \begin{picture}(400,200)
   \put(40,100){\vector(1,0){320}}
   \put(200,20){\vector(0,1){180}}
   \put(188,88){$0$}
   \put(365,98){$x$}
   \put(198,205){$V(x)$}

   \qbezier(50,105)(80,115)(100,145)
   \qbezier(100,145)(150,220)(150,100)
   \put(150.5,100){\line(0,-1){50}}
   \put(150.5,50){\line(1,0){100}}
   \put(250.5,50){\line(0,1){50}}
   \qbezier(250,100)(250,220)(300,145)
   \qbezier(300,145)(320,115)(350,105)
  \end{picture}
\end{center}
\caption[]{An example of attractive potential 
with complex energy eigenvalues. }
\end{figure}
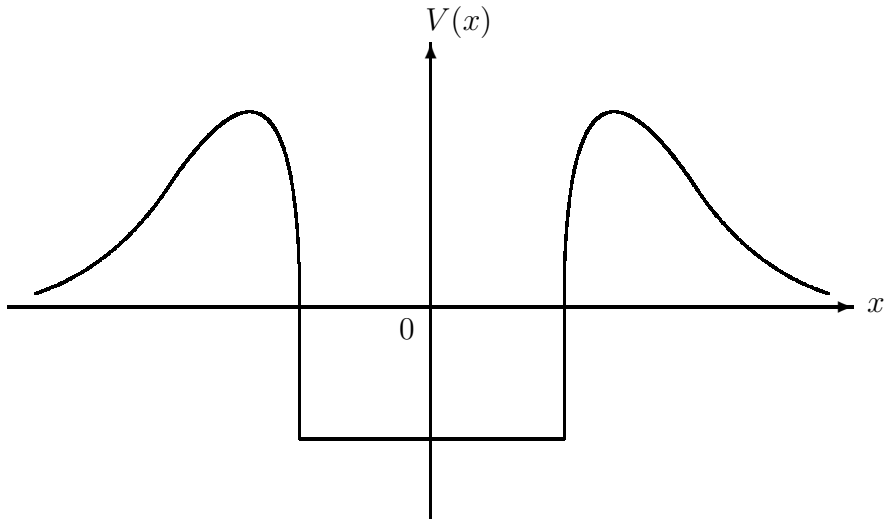
For instance, electric potentials of atoms for incoming electrons 
will be approximated by such potentials. 
Namely there is an attractive force near the center 
arising from the positive charge of nucleus, which is 
surrounded by the repulsive force induced from the electrons trapped 
in the atom. 
There will be a possibility for interpreting chemical bonds in this 
scheme. 

It is also pointed out that these quasi-stable states are connected by flows 
which are generally observable in quantum mechanics. 
This property indicates that those states can have some classical properties. 
It seems to be very attractive to describe phenomena being in borders 
between quantum processes and classical ones such as mesoscopic processes 
in terms of the present scheme. 
We should notice that the idea of the connection by flows are applicable 
to real energy eigenvalues states contained in usual Hilbert spaces. 
\vskip5pt

Finally we would like to comment on cases where imaginary eigenvalues are 
not independent of real ones. 
In general there may be some relations between real and imaginary 
energy eigenvalues ($\E_j$ and $\G_j$). 
In such cases the formula (4) for the number of combinations 
should be written as 
\EQ
W({\cal E})=
\sum_{\E_1}\cdots\sum_{\E_N}\D_{\sum_j\E_j,E}\  W^\Im (\GA:\E_1,...,\E_N),
\EN 
where 
$\D_{\sum_j\E_j,E}$ is the Kronecker delta symbol representing the relation 
$\sum_j\E_j=E$ and 
$W^\Im (\GA:\E_1,...,\E_N)$ stands for the sum over states 
with respect to the freedom of the imaginary energies 
when the real energy eigenvalues of all constituents are fixed. 
It is obvious that the above formula turns to that of (4), when 
$W^\Im (\GA:\E_1,...,\E_N)$ does not depend on the real 
energy eigenvalues. 
We can follow the same procedure in deriving the entropy. 
It is, however, very hard to obtain a general formula 
for the entropy, 
because the situations so much depend on the relations between 
the real and imaginary eigenvalues. 
We have to investigate very carefully and precisely the relation for 
each interaction, that is, we have to solve Gel'fand-triplet problems 
for the interaction. 
It is, however, stressed that 
the existence of quasi-stable states composed of complex energy 
eigenstates is guaranteed because of pairings of complex energy 
eigenvalues, when states having non-zero imaginary-energies are involved 
in Gel'fand-triplet solutions. 
We should also investigate the possibility of entropy transfer between 
two entropies $S^\Re$ and $S^\Im$ given in (5). 
\pagebreak

\hfil\break
{\bf References}
\hfil\break
[1] A. Bohm and M. Gadella, {\it Dirac Kets, Gamow Vectors and Gel'fand 
Triplets}, Lecture Notes in Physics, Vol.~348 (Springer, Berlin) 1989.
\hfil\break
[2] T. Shimbori and T. Kobayashi, Il Nuovo Cimento B (to appear, 2000).
\hfil\break
[3] T. Shimbori, {\it Operator Methods of the Parabolic Potential Barrier}, 
UTHEP-415, quant-ph/9912073.
\hfil\break
[4] T. Shimbori and T. Kobayashi, {\it Stationary Flows of 
Parabolic Potential Barrier in Two Dimensions} (to appear, 2000).
\hfil\break
[5] T. Shimbori and T. Kobayashi, {\it ``Velocities'' in Quantum Mechanics}, 
UTHEP-422, quant-ph/0004086.
\hfil\break
[6] For instance, see the following textbook. L. D. Landau and E. M. Lifshitz, 
{\it Quantum Mechanics} 3rd ed. (Pergamon, Oxford, 1977).







\end{document}